\documentclass{iau}
\usepackage{graphicx}

\def\apss{Ap\&SS}
\def\aap{A\&A}

\def\aaps{A\&AS}

\def\mnras{MNRAS}

\def\pasp{PASP}

\def\arcsec{\hbox{$^{\prime\prime}$}}

\title[Orientation of Galactic Bulge PNe] %% give here short title %%
{Orientation of Galactic Bulge Planetary Nebulae 
toward the Galactic Center}

\author[Ashkbiz Danehkar \& Quentin A. Parker]   %% give here short author list %%
{Ashkbiz Danehkar$^1$
 \and Quentin A. Parker$^{1,2}$}

\affiliation{$^1$Department of Physics and Astronomy, Macquarie University, Sydney, NSW 2109, Australia  \\[\affilskip]
$^2$Australian Astronomical Observatory, PO Box 915, North Ryde, NSW 1670, Australia }

\pubyear{2015}
\volume{312}  %% insert here IAU Symposium No.
\pagerange{119--126}
\jname{Star Clusters and Black Holes in Galaxies across Cosmic Time}
\editors{R. Spurzem \& F. Liu, eds.}
\begin{document}

\maketitle

\begin{abstract}
We have used the Wide Field Spectrograph on the Australian National University 2.3-m telescope to perform the integral field spectroscopy for a sample of the Galactic planetary nebulae. The spatially resolved velocity distributions of the H$\alpha$ emission line were used to determine the kinematic features and nebular orientations. Our findings show that some bulge planetary nebulae toward the Galactic center have a particular orientation. 
\end{abstract}

The majority of planetary nebulae (PNe) show predominantly axisymmetric morphologies, i.e. elliptical and bipolar.
There is a strong argument in favor of the axisymmetric morphologies of PNe being related to the angular momentum of their central engines (e.g., \cite[Miszalski et al. 2009]{Miszalski2009b}; 
\cite[Nordhaus et al. 2010]{Nordhaus2010}). Recently, it has been found that the nebular inclinations are closely related to the orbital inclinations of their close-binary central stars (see e.g., \cite[Jones et al. 2012]{Jones2012}; \cite[Tyndall et al. 2012]{2012}).

PN as a test particle can be used to trace the dynamics of the Galaxy. The major axes of bipolar PNe are assumed to be perfectly aligned with the angular momentum vectors of their central stars, so a population of bipolar Galactic PNe can be used to determine the angular momentum distribution of the local stars within the Galaxy. Recent studies showed that some bipolar PNe within the bulge near the Galactic center have a homogeneous orientation (\cite[Weidmann \& D{\'{\i}}az 2008]{Weidmann2008}; \cite[Rees \& Zijlstra 2013]{Rees2013}). 

\begin{figure}[b]
\begin{center}
 \includegraphics[width=1.2in]{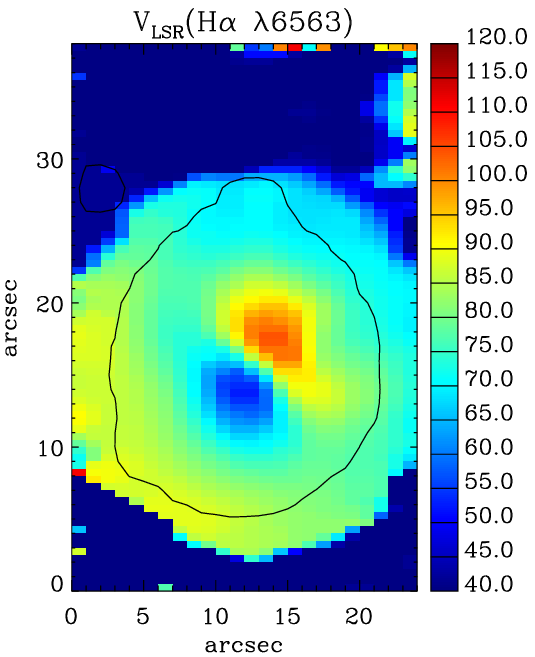}%
 \includegraphics[width=1.2in]{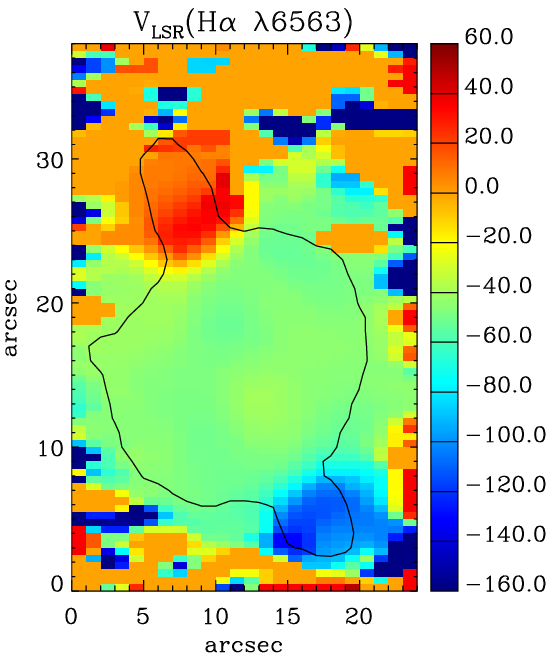}%
 \includegraphics[width=1.2in]{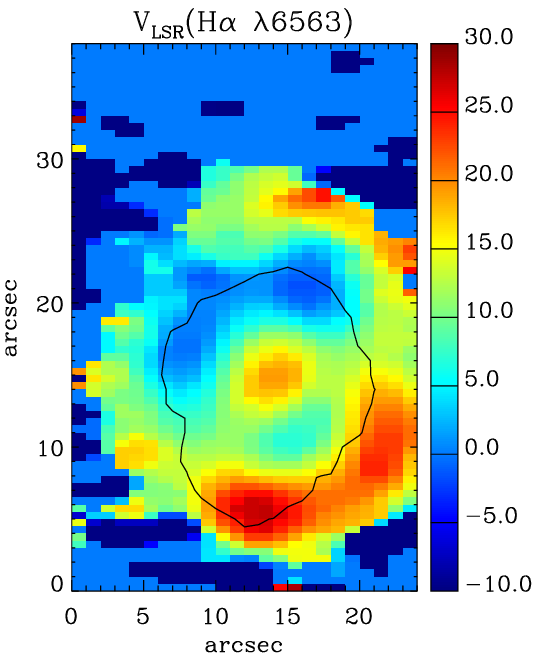}%
 \includegraphics[width=1.2in]{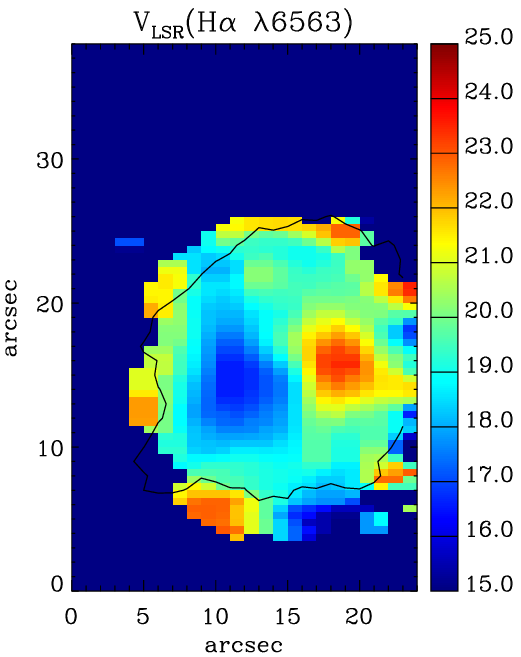}%
 \includegraphics[width=1.2in]{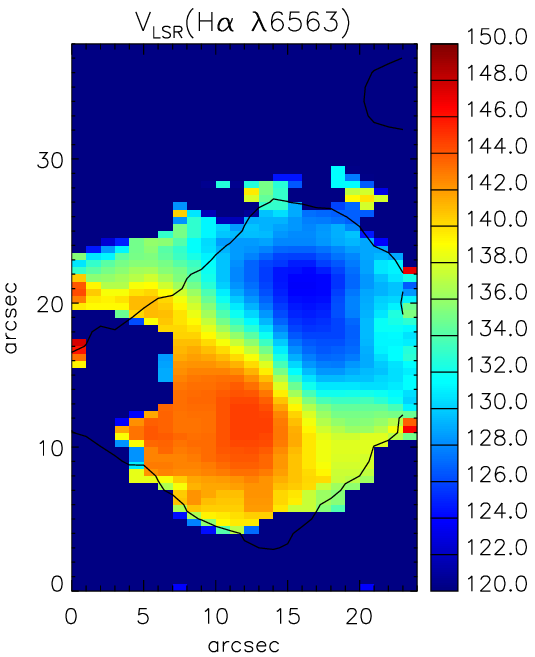} 
 \includegraphics[width=4.5in]{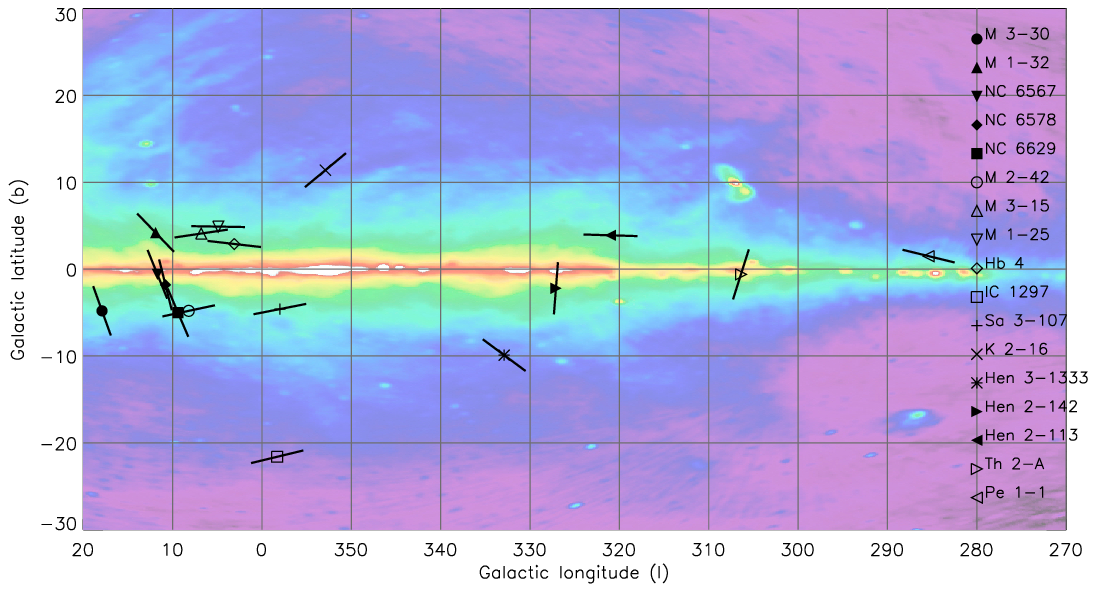} 
 \caption{Top panels: From left to right, IFU kinematic maps of M\,3-30, Hb\,4, IC\,1297, NGC\,6578, and NGC\,6567, respectively. North is up and east is toward the left-hand side. 
Bottom panel: Orientations of PNe in Galactic coordinates. The lines display the nebular symmetric axes. The background is the radio continuum (408 MHz) brightness from \cite[Haslam et al. (1982)]{Haslam1982}.   \label{fig1}}
\end{center}
\end{figure}

\subsection{IFU Observations of Galactic Planetary Nebulae}

In this study, we aim to disentangle 3D gaseous structures of some Galactic PNe by means of integral field spectroscopy. We used the Wide Field Spectrograph (\cite[WiFeS; Dopita et al. 2010]{Dopita2010}) on the 2.3-m ANU telescope at Siding Spring Observatory. WiFeS is an image-slicing integral field unit (IFU) feeding a double-beam spectrograph. It has a field-of-view of $25\arcsec\times38\arcsec$  and a spatial resolution of  $1\arcsec$. Our observations were carried out with a spectral resolution of $R\sim 7000$. The spectra were reduced using the IRAF pipeline WIFES (\cite[described in detail by Danehkar et al. 2013]{Danehkar2013a}).

We obtain the kinematic features by applying a Gaussian curve fit to each spaxel of IFU datacube (see Fig. \ref{fig1}). The observed kinematic maps are then compared to 3-D kinematic models in order to find the orientations of the nebular symmetric axes onto the plane of the sky, i.e. the position angle (PA) measured from the north toward the east in the equatorial coordinate system (\cite[see Danehkar et al. 2014]{Danehkar2014}). The PA is transferred into the Galactic position angle (GPA) in the Galactic coordinate system. Fig. \ref{fig1} (bottom) shows orientations of the PNe based on their symmetric axes projected onto the sky plane.

The spatial orientations of the Galactic PNe have been studied by few authors (\cite[Phillips 1997]{Phillips1997}; \cite[Corradi et al. 1998]{Corradi1998}; \cite[Weidmann \& D{\'{\i}}az 2008]{Weidmann2008}; \cite[Rees \& Zijlstra 2013]{Rees2013}). Although the orientations of nearby Galactic PNe are found to be completely arbitrary in the Galactic disc, some distant PNe seem to have a particular orientation within the Galactic bulge, as seen in Fig. \ref{fig1} (in agreement with \cite[Weidmann \& D{\'{\i}}az 2008]{Weidmann2008}; \cite[Rees \& Zijlstra 2013]{Rees2013}). Galactic bulge PNe are affected by the bugle dynamics and the supermassive black hole located at the Galactic center, and are required to be observed and investigated further.

\section*{Acknowledgments}
AD acknowledges travel grants from IAU and Australian Institute of Physics.

\end{document}